\documentclass[12pt]{article}
\usepackage{amsmath, amsfonts, amscd, euscript}

\newcommand{\be}[1]{\begin{equation}\label{#1}}

\newcommand{\dd}{{d}}

\newcommand{\ee}{\end{equation}}

\newcommand{\amg}{\vec{\Omega}}

\newcommand{\R}{\mathbb{R}}
\newcommand{\com}{\mathbb{C}}

\newcommand{\loc}{\mathfrak{L}}

\newcommand{\Z}{\mathbb{Z}}

\newcommand{\latt}{\mathcal{U}\mathcal{P}}
\newcommand{\tropos}{\mathcal{T}\!\!{\rm r}_{CH}}
\title{\LARGE\bf Sheafifying Consistent Histories}

\author{Ioannis Raptis\thanks{Theoretical
Physics Group, Blackett Laboratory of Physics, Imperial College of
Science, Technology and Medicine, Prince Consort Road, South
Kensington, London SW7 2BZ, UK; e-mail: i.raptis@ic.ac.uk}}

\date{}

\begin{document}

\maketitle

\begin{abstract}

\noindent\footnotesize Isham's topos-theoretic perspective on the
logic of the consistent-histories theory \cite{isham97} is
extended in two ways. First, the presheaves of consistent sets of
history propositions in their corresponding topos originally
proposed in \cite{isham97} are endowed with a Vietoris-type of
topology and subsequently they are sheafified with respect to it.
The category resulting from this sheafification procedure is the
topos of sheaves of sets varying continuously over the
Vietoris-topologized base poset category of Boolean subalgebras of
the universal orthoalgebra $\latt$ of quantum history
propositions. The second extension of the topos in \cite{isham97}
consists in endowing the stalks of the aforementioned sheaves,
which were originally inhabited by structureless sets, with
further algebraic structure that also enjoys a quantum causal
interpretation {\it \`a la}
\cite{rap1,malrap,rap3,rapzap1,rapzap2} so as to arrive at the
topos of consistent-histories of quantum causal sets. Not being
able to resist the temptation, we speculate on a possible
application of such topos-theoretic models to the problem of
quantum gravity ({\it ie}, when spacetime structure, causality and
its dynamics are supposed to be treated quantum mechanically)---an
application that has been anticipated on general grounds by
Butterfield and Isham \cite{buttish4} and partly worked out in a
special finitary algebraic, sheaf-theoretic and categorical
setting by this author \cite{rap1,rap2,malrap,rap3,ces}. In
particular, rather general, but striking, similarities between the
topos of consistent-histories of quantum causal sets that arises
from our second extension of \cite{isham97}, the topos of finitary
spacetime sheaves of non-abelian incidence algebras modeling a
dynamical and locally finite quantum causality and its associated
non-commutative topology in \cite{malrap,rap3,ces}, as well as the
recently proposed quantum spacetime scenario based on the
so-called quantum causal histories of Markopoulou \cite{marko},
are exposed. The paper closes with this author's personal views,
anticipations and speculations about the future of the general
research program of applying sheaf and topos-theoretic ideas
primarily to quantum gravity and then to quantum logic.

\end{abstract}

\section{\Large Introduction cum Motivation}\label{intro}

The consistent-histories (CH) approach to quantum theory
\cite{griffs,omnes,gelhar1} presents a sound alternative
interpretation of quantum mechanics---one that is less
instrumentalist or operational and philosophically more realist
than the standard `Copenhagen' one, for it purports to avoid
altogether the notorious Heisenberg {\it schnitt} of the usual
theory. Perhaps, such an approach is more fit to support a quantum
theoresis of the universe as a whole ({\it ie}, quantum cosmology)
\cite{gelhar2,har} where an observer-system split appears to be
highly inappropriate and {\itshape prima facie} meaningless.
Certainly, one expects that ideas that were born out of the CH
version of quantum theory should apply straightforwardly to our
quite general endeavor of applying quantum mechanical concepts,
results and techniques to the structure and dynamics of spacetime
\cite{isham97,buttish4}, as well as to the associated problem of
quantizing causality \cite{marko}; altogether, to the general and
by now quite broad and diverse quantum gravity research program.

The quantum sort of logic that uderlies the CH
theory\footnote{Hereafter we will refer to this logic as `quantal
logic' in order to distinguish it from the quantum logic proper
that underlies the standard quantum mechanics \cite{birkhoff}.}
has been beautifully exposed in \cite{isham94}. Subsequently, this
quantal logic was subjected to a topos-theoretic analysis
\cite{isham97} which revealed the theory's strong `neorealist'
undertones in the following sense: the universal orthoalgebra
$\latt$ of history propositions admits non-trivial localizations
or `contextualizations' (of truth) over its classical Boolean
subalgebras. More technically speaking, it was shown that one
cannot meaningfully assign truth or semantic values to propositions about
histories globally in $\latt$, but that one can only do so
locally, that is to say, when the propositions live in certain
Boolean sublattices of $\latt$---the classical sites, or windows
\cite{buttish1,buttish2,buttish3}, or even points
\cite{rap3,mul1,mul2} within the quantum lattice
$\latt$\footnote{Here we will use the names `orthoalgebra' and
`lattice' interchangeably for $\latt$, although strictly speaking
the latter is a stronger algebraic structure than the former
\cite{foulis}.}. Moreover, the simultaneous consideration of {\em
all} such Boolean subalgebras and {\em all} consistent sets of
history propositions\footnote{Isham's assumption of {\em all}
consistent sets of history propositions may be called ``the
principle of histories' democracy". See next paragraph.} leads one
to realize that the `internal logic' of the CH theory is neither
classical (Boolean) nor quantum proper, but
intuitionistic\footnote{As mentioned earlier, Isham in
\cite{isham97} uses the epithet `neorealist' for the quantal logic
of the CH theory in its topos-theoretic guise. Quite resonably, we feel, 
one could also coin this logic `neoclassical' 
\cite{rap3}---this name referring to the departure of the Brouwerian 
logic of the topos of consistent-histories in \cite{isham97} from the two-valued Boolean lattice 
calculus 
obeyed by the states of a classical mechanical system which are modeled 
after point subsets of its 
phase space. See also the next section for more about this significant departure of the quantal 
logic of the CH theory from classical Boolean logic.}. 
This result befits the fact that the relevant
mathematical structure involved in \cite{isham97}, namely, the
collection of presheaves of sets varying over the poset category
of Boolean sublattices of $\latt$, is an example of an abstract
mathematical structure known as a topos
\cite{lawv,bell,maclane,asperti}, for it is a general result in
category theory that every topos has an internal logic that is
strongly typed and intuitionistic
\cite{gold,lambek,sel91,maclane}. This result also seems to suit
the primitive intuition that some kind of `many-world-views'
\cite{isham97}, or to the same effect,
`all-consistent-histories-view' of the logic of the CH scheme will
point to the appropriate semantics of the theory, since a modal
Kripke-type of `possible-worlds' semantics \cite{kripke} has been
found to apply to a very similar topos of preasheaves of variable
sets structure underlying the non-distributive quantum logic
proper \cite{rawse}\footnote{As it is also emphasized in
\cite{isham97}, the (locally) intuitionistic quantal logic of the
CH theory, while nonclassical ({\it ie}, non-Boolean) in the sense
that the law of excluded middle ({\it tertium non datur}) does not
hold in it, it still is distributive (at least locally; see
below), in contradistinction to quantum logic proper which at a
`global' level ({\it ie}, when the focus is not restricted solely
on propositions dwelling in the logic's Boolean sublattices or
classical points) appears to obey a prominently non-distributive lattice
calculus \cite{birkhoff}, although it too may be shown to be
locally intuitionistic and neorealist when viewed from a
topos-theoretic perspective
\cite{buttish1,buttish2,buttish3,rawse}.}, as it too was nicely
revealed by Isham {\it et al.} in the trilogy 
\cite{buttish1,buttish2,buttish3}\footnote{We will return to
discuss this trilogy in more detail in the last section.}.

The reader must have noticed by now that no allusion to the
measure or probability-theoretic attributes of the CH theory has
been made so far. From a physical point of view, one of the most
tantalizing questions one can raise about the CH theory is what
singles out or `realizes' a complete set of history propositions
as the `actual' or `real' one from all other such
sets\footnote{For the notion of completeness of a set of history
propositions, see \cite{isham97}.}. In the first place, it was the
consideration of {\em all} the sets of history propositions that
are consistent relative to a $\com$-valued measure $\dd$---the
so-called `decoherence functional'\footnote{Again, see
\cite{isham97} for a brief discussion about this object and the
$\dd$-consistent sets of history propositions associated with
it.}---that led Isham \cite{isham97} to question whether one
should restrict probability assignments solely on propositions
about $\dd$-consistent sets rather than, say, consider the larger
ensemble of history propositions that are complete, but not
necessarily consistent relative to a decoherence functional.
Isham's challenging of $\dd$-consistency paid off since he
postulated the aforementioned principle of histories'
democracy\footnote{That is to say, to assume {\it ab initio} {\em
all} $\dd$-consistent sets, rather than single out by hand a
preferred or actual one.} which guided him rather
straightforwardly to the notion of sieves of $\dd$-consistent
coarse-grainings of complete sets of history propositions, then to
their associated valuation presheaves and ultimately to the
neorealist topos thereof.

Similarly, in the present paper we are not going to occupy
ourselves with probabilistic features of the CH theory, for Isham's
topos-theoretic scheme seems to us to be convincing, rich and
compelling enough. Rather, we are going to attempt to extend his
work \cite{isham97} in two fronts, which may be coined ($i$) `the
base front', and ($ii$) `the stalk front', for reasons to be
explained below:

\begin{itemize}

\item ($i$) \underline{\bf The base front}: The first extension of
Isham's paper \cite{isham97} consists in an attempt to endow the
presheaves of sets of consistent-history propositions with a
suitable topology. This `topologization of histories' will lead us
effortlessly to sheafifying their respective presheaves, thus
convert the topos organization of the latter to the topos of
sheaves of {\em continuously} variable sets over the base poset
category of Boolean subalgebras of $\latt$ now regarded as a
background topological substratum proper. Such a move, apart from
its mathematical naturalness\footnote{That is to say, it is
customary in mathematics when a presheaf-like structure appears
during the development of a theory, that the next rather natural
question that one may ask is whether the base space admits a
topology; hence, whether the presheaf can be converted to a sheaf
\cite{bredon,maclane,mallios,rap2}.}, is expected to unveil
otherwise concealed (by the conventional non-topos-theoretic
histories formalism) topological features of the CH theory. In any
case, from a purely topos-theoretic perspective on the CH theory,
such a move appears to be all the more legitimate, because the
collection of sheaves (of algebraic structures of any kind) over a
locale---the most general sort of a topological space
\cite{maclane}---appears to be the most canonical paradigm of a
topos \cite{bor1,bor2,maclane}. At least, having topologized the
base space over which the overlying objects\footnote{In
\cite{isham97} these objects are structureless sets.} are varying,
we are able to qualify Lawvere's adverb `continuously' in
\cite{lawv}. In any case, it would be nice to have a
consistent-histories analogue of the topos $\mathbf{Sh}(X)$ of
sheaves of sets varying over a continuous spacetime manifold
$X$---the mathematical universe in which arguably all quantum,
albeit flat, field theories have been hitherto formulated
\cite{sel91}\footnote{This will also prompt us to look for a
`curvaceous' topos, since the CH theory purports to address the
problem of quantum gravity \cite{isham97,buttish4,rap3,ces}. See
also ($ii$) next as well as section \ref{algebran}.}.

\item ($ii$) \underline{\bf The stalk front}: The topologization and
concomitant sheafification of the presheaves in ($i$) will be
followed by an algebraization of the stalks of the resulting
sheaves. That is to say, instead of considering only structureless
sets as inhabiting the stalks of the resulting sheaves as in
\cite{isham97}, we will assume that the latter are occupied by
so-called {\em quantum causal sets} which are finitary algebraic
structures \cite{rap1,malrap,rap3,ces,rapzap1,rapzap2}. The
ultimate hope of such a move is to be able to view the resulting
topos of continuously variable quantum causal sets as the proper
mathematical universe in which to study the dynamical variations
of quantum causality---a dynamics that is expected to be at the
heart of yet another algebraic approach to quantum gravity
\cite{rap1,rap2,malrap,rap3,ces,rapzap1,rapzap2}. We will also see
that this second stalk-wise extension of Isham's topos-theoretic
perspective on the CH theory, together with its quantum causal
interpretation, is very similar to Markopoulou's recent quantum
causal histories scenario for quantum spacetime structure and
gravity \cite{marko} which purports to be a successful fusion of
general ideas from the CH theory with ideas from another quite
promising finitistic-causal approach to quantum gravity coined
`causal set theory' which has been around for more than a decade
now \cite{bomb87,sor1,sor2,sor4,sor5}. Certainly, a bonus from
soldering algebraic quantum structures of significant operational
character, like the quantum causal sets in
\cite{rap1,malrap,rap3,rapzap1,rapzap2}, on consistent-histories
is that it enables us to reinstate to a certain extent some sort
of operationality (if not strict instrumentalism!) in the CH
theory---a theory whose interpretational philosophy at first sight
appears to have a purely realist flavor \cite{isham97}\footnote{So
that, for instance, even probabilities are interpreted as
propensities in the CH theory, and in a strong sense history
propositions are about the universe `as such' or `in itself'
\cite{isham97}. This seems to tie well with the aforementioned
existence of an `internal logic' for every topos, hence it further
justifies Isham's fundamental insight of assuming a topos
perspective on the CH theory.}.

\end{itemize}

\medskip

The present paper is organized as follows: in the next section we
give a brief review of the neorealist consistent-histories topos
constructed in \cite{isham97} and we highlight its features that
are of relevance to our labors in the subsequent sections. In
section \ref{sheafn} we topologize the base poset category $\mathcal{B}$ 
of Boolean subalgebras of $\latt$ over which sets
are assumed to vary in \cite{isham97} by endowing history 
propositions with a Vietoris-type of topology, and then we sheafify the
associated presheaves of sets in their topos 
organization relative to the locale of open subsets of history propositions in 
the Vietoris-like topological space that they constitute. In section 
\ref{algebran} we algebraize the stalks
of sheaves that resulted from the topologization-sheafification
procedure of the previous section by assuming that these fibers
are inhabited by finitary algebraic quantum causal sets
\cite{rap1,malrap,rap3,rapzap1,rapzap2}, rather than merely by
structureless sets as in \cite{isham97}. Thus we arrive at the
{\em topos of consistent-histories of quantum causal sets}
(QCHT)\footnote{Initials for `Quantum Causal Histories Topos'.}
and compare it with the quantum causal histories scenario proposed
by Markopoulou in \cite{marko}. We also compare the QCHT with 
certain sheaf-theoretic models, of a finitary, causal 
and quantal flavor, for the kinematics of  
Lorentzian quantum gravity suggested in \cite{malrap},
 as well as with some related algebraic 
attempts of this author to  
arrive at a cogent non-commutative or `quantum' topology, of a strong
finitistic and causal flavor, for quantum gravity
in \cite{rap3,ces}. In the concluding section we discuss
various formal, but rather impressive, similarities
between our QCHT and another topos-like structure that has 
recently appeared in connection with the famous Kochen-Specker
`no-go' theorem (or paradox!) of quantum logic proper 
\cite{buttish1,buttish2,buttish3}. 
The paper closes with some of this author's
personal and undoubtedly subjective views about the future course
of development of the sheaf and topos-theoretic approach to the CH
theory in particular and to the broader quantum gravity research program in
general. More specifically, the possibility of infusing some differential 
geometric ideas 
and constructions to the CH theory by sheaf and topos-theoretic means, 
ultimately with 
an eye towards applying the resulting structures to quantum gravity, is 
briefly entertained at the end.

\section{\Large\bf Isham's neorealist consistent-history topos revisited}\label{neotopos}

In this section we recall briefly concepts and results from the
topos-theoretic perspective  on the logic of the CH theory assumed
in \cite{isham97} that are going to be of relevance to the rest of
the paper. The reader should refer to Isham's original paper for a
more thorough analysis of these elements.

The first element of structure of the quantal logic of the CH
theory is that its propositions (about histories) form an
orthoalgebra $\latt$ \cite{isham94}. A representation  of $\latt$
by projection operators in a suitable tensor product Hilbert space
${\cal{H}}$ instantly reveals its `inherently quantum' nature in
the sense that the resulting projection lattice
${\cal{L}}({\cal{H}})$ is characteristically non-distributive
\cite{isham94}---the quintessential feature of quantum logic
proper \cite{birkhoff}. A question that might occur to a quantum
logician who is familiar with the logic of the CH theory and who
is of a strong philosophical or `toposophical' bent\footnote{It is
not uncommon in mathematics' social jargon for a categorist or
topos-theorist who is interested in wider applications or
philosophical extensions of topos theory to be called a
`toposopher'.} is whether, apart from the unified formal
mathematical or `syntactic' structure that underlies both the
quantal logic of the CH theory and the usual quantum logic of the
conventional quantum mechanics ({\it ie}, the non-distributive
orthomodular lattice calculus \cite{rawse}), there are other
deeper similarities, or perhaps more importantly, differences
between the two logics. For instance, one may enquire whether the
valuation or `truth-theoretic' and other associated `semantic'
aspects of the two schemes are also analogous to or fundamentally
different from each other, and whether a topos-theoretic stance
against these theories will shed light on such a
comparison\footnote{The reader must await section \ref{affinities}
for a brief comparison between the quantal and quantum logics from
a topos-theoretic viewpoint.}. After all, in spite of their
impressive formal mathematical analogies at the syntactic
proposition lattice level, the physical semantics or philosophical
interpretation of the two theories are significantly different, as
it was briefly mentioned in the introduction.

It is fair to say that Isham's assumption of a topos-theoretic
perspective on the CH theory  was predominantly motivated by an
interest to explore more `qualitative' truth-theoretic, semantic
or `valuational' aspects of the quantal logic of
consistent-histories, although there were also other
`quantitative' probabilistic or measure-theoretic aspects of the
CH theory that appealed to him originally. The latter, however, we
are only loosely going to address here. Below we summarize the
basic ideas and results from \cite{isham97} by itemizing
them\footnote{The reader is assured that sophisticated
topos-theoretic jargon and highly technical concepts or intricate
results from topos theory will be seldom used in this physically,
rather than mathematically, oriented paper. When a technical
concept is mentioned, or when a theorem and result is quoted,
references to the relevant mathematics literature, rather than an
analytical discussion, will be given.}:

\begin{itemize}

\item (a) \underline{\bf Presheaves:} Presheaves\footnote{For the technical notion of presheaves,
consult \cite{bredon,maclane,mallios,rap2}. We will discuss them
more analytically in section \ref{sheafn}.} arise in
\cite{isham97} from considering the notion of {\em varying} or
{\em variable sets} \cite{lawv} in the context of the CH theory.
In particular, of central importance in \cite{isham97} is the
notion of presheaves of sets over the poset category
${\mathcal{B}}$ of Boolean subalgebras of
$\latt$\footnote{${\mathcal{B}}$ is the collection of Boolean
sublattices of $\latt$ ordered by set-theoretic inclusion
$\subseteq$ which may be interpreted as `coarse-graining of
histories' in the sense that $W_{1}\subseteq W_{2}$ reads `$W_{1}$
is coarser than $W_{2}$' or equivalently that `$W_{2}$ is finer
than $W_{1}$'.}, denoted by ${\rm Set}^{\mathcal{B}}$.
${\mathcal{B}}$ provides an abstract `temporal' background (base)
domain (space) over which sets (or algebraic structures of any
kind\footnote{For presheaves of algebraic structures more
elaborate than sets, see section \ref{algebran}.}) are supposed to
vary---its abstract character consisting in our identifying the
notion of `temporal order' or `succession' with the process of
coarse-graining of consistent-histories\footnote{The use of a
poset as a `temporal support' or as a `general domain of variation
of a causal nature' is also used in \cite{malrap,rap3,marko}. In
connection with presheaves of sets varying in time \cite{lawv},
the coarse-graining relation defining the base poset category
$\mathcal{B}$ had an analogous connotation for Isham in 
\cite{isham97}. Again, we will encounter such structures in
section \ref{algebran}.}. The structures inhabiting the presheaves
are seen as generalized `truth spaces'---realms in which truth
assignments or `valuation functions'\footnote{For the technical
notion of valuations, refer to \cite{isham97}.} on the history
propositions in $\latt$ take their values. As a result, and from a
geometrical perspective, the Boolean subalgebras $W$ of $\latt$ in
$\mathcal B$ may be viewed as the `classical localization sites'
or `points of contextualization of truth', or even as `classical
windows of access to the quantum system's states'
\cite{isham97,buttish1,buttish2,buttish3,rawse,rap3}, within the
quantum space $\latt$. The interpretation for them that we favor
here is a more `temporal' one as `local stages of truth'
\cite{isham97}---frozen instances or `snapshots' of truth value
assignments on compatible history propositions living in each
$W$\footnote{The classical Boolean windows $W$ in $\latt$ are
generated by propositions whose corresponding projection operators
on the (closed) subspaces of ${\mathcal{L}}({\mathcal{H}})$
commute with each other, hence the epithet `compatible'
\cite{isham94,isham97,rawse}.}---in the `global flow of truth over
the partially ordered support $\mathcal B$'. Let us call the
presheaves of the form ${\rm Set}^{\mathcal{B}}$ `the valuation
presheaves associated with $\latt$'.

\item (b) \underline{\bf Sieves:} Sieves\footnote{For the technical
definition of sieves, consult \cite{maclane}.}
arise in \cite{isham97} in close connection with the presheaves
${\rm Set}^{\mathcal{B}}$. Isham naturally arrived at sieves by
questioning whether only so-called second-level propositions about
histories relative to a decoherence functional
$\dd$\footnote{Rougly, a second-level proposition about a history
is of the form `history $a$ is realized with probability $p$
relative to a chosen decoherence functional $\dd$'; hence,
second-level history propositions are about $d$-consistent sets of
histories in $\latt$ where the usual Kolmogorov axioms of
probability theory appear to apply rather naturally \cite{isham97}.} should be
considered in probabilistic predictions about histories in the CH
theory. In effect, he noticed that by coarse-graining a complete
set $C$ of history propositions that {\em is not} $\dd$-consistent
one could obtain a set $C'$ that {\em is}; moreover, any further
coarsenings of $C'$ still yield $\dd$-consistent sets---the two
defining properties of a sieve structure on a poset such as
$\mathcal B$. We may call these sieves `the coarse-graining sieves
on $\mathcal B$'. An even more suggestive result from
\cite{isham97} is that for every object $W_{0}$ in the poset
category $\cal B$ ({\it ie}, at every stage during the `unfolding
of truth' in $\latt$) the collection of {\em all} coarse-graining
sieves based or soldered at $W_{0}$\footnote{Thus effectively the
consideration of {\em all} complete sets of history propositions
in $\latt$ (histories' democracy).} form a logico-algebraic
structure isomorphic to a Heyting algebra\footnote{See
\cite{gold,lambek,bell,maclane} for a definition of this lattice
structure.} which is supposed to encode the lattice calculus of
intuitionistic logic \cite{gold,lambek,bell,maclane}. Let us
symbolize this object by $\Omega(W_{0})$. This discussion brings
us to the crucial fact about the collection of all presheaves
${\rm Set}^{{\mathcal{B}}}$ of sets varying over $\mathcal B$.

\item (c) \underline{\bf The topos of presheaves:} The category of all
objects of the form ${\rm Set}^{{\mathcal{B}}}$
and presheaf morphisms\footnote{For the notion of (pre)sheaf
morphisms, refer to \cite{bredon,maclane,mallios}.} between them
is an example of a topos
\cite{gold,lambek,bell,sel91,asperti,maclane}, which we may
symbolized as ${\mathcal{T}}_{CH}$\footnote{The
consistent-histories topos.}. The aforementioned Heyting algebra
object $\Omega$ in ${\mathcal{T}}_{CH}$ is known as the topos'
subobject classifier\footnote{See
\cite{gold,lambek,bell,sel91,asperti,maclane} for a description of
this object.}. By interpreting the latter as a generalized truth
or semantic value space, Isham interpreted the lattice morphisms
object-wise (or window-wise) in ${\mathcal{T}}_{CH}$ of the form

\begin{equation}\label{e1}
\mathfrak{V}:\, W\rightarrow\Omega(W), \, (\forall W\in{\mathcal{B}})
\end{equation}

\noindent as localized (or contextualized\footnote{The Boolean
windows in ${\mathcal{B}}$ providing the localization sites or
contexts.}) valuations or truth value assignments to
(second-level) history propositions in $\latt$. In fact, this is a
`corollary' of the following theorem: the presheaves in the
consistent-histories topos ${\cal{T}}_{CH}$ admit no global
sections\footnote{For the technical notion of sections of
(pre)sheaves, see \cite{maclane,mallios}.} over $\latt$; they only
do so locally, that is, when restricted over the Boolean
sublattices of $\latt$\footnote{That local sections of the
presheaves give rise to their subobjects in ${\cal{T}}_{CH}$,
hence to presheaf morphisms (valuations) {\it \`a la} (\ref{e1}),
is a well known fact in topos theory
\cite{asperti,maclane,rawse}.}. We may resume this by saying that
`in the quantal logic of the CH theory truth is localized or
contextualized on the classical Boolean subalgebras of the
universal quantum proposition lattice $\latt$'---which discussion
brings us to an even more `universal' result in topos theory.

\item (d) \underline{\bf The internal language of the topos
$\mathbf{{\mathcal{T}}_{CH}}$:} The internal language or logic
of the consistent-histories topos ${\cal{T}}_{CH}$ is
intuitionistic type theory \cite{gold,lambek,bell,sel91,maclane}.
This is effectively encoded in the subobject classifier $\Omega$
of ${\cal{T}}_{CH}$ which, as noted above, is a Heyting
algebra---the logic algebra of intuitionism\footnote{As noted in
the introduction, the main characteristic of the Heyting calculus
of intuitionistic logic is that double negation of a proposition
is not its assertion, which reflects a violation of the law of
excluded middle of the two-valued Boolean logic which has a
unipotent negation unary operation.}. This is in striking
asymphony with the logic of the topos $\mathbf{Set}$ of sets whose
subobject classifier is the Boolean binary alternative ${\mathbf
2}$\footnote{The trivial Boolean algebra $\{0,1\}$ consisting of
the truth values $0$ (false) and $1$ (true).}, hence whose
internal logic is inherently Boolean. Due to this difference in
logic, $\mathbf{Set}$ is thought of as a universe of constant or
`frozen' (perhaps in time \cite{lawv}) sets, while
${\mathcal{T}}_{CH}$ may be thought of as a realm of variable sets
\cite{lawv,sel91,isham97}. Furthermore, it is precisely due to
this generalization of the Boolean binary alternative of the
classical logic of constant sets in $\mathbf{Set}$ to the Heyting
algebra subobject classifier of the intuitionistic logic of
variable sets in ${\mathcal{T}}_{CH}$ that Isham coined the latter
topos `neorealist' in \cite{isham97} and, in a slightly different
context, this author `neoclassical' in \cite{rap3}. Neorealism
then pertains precisely to the localization or contextualization
of truth value assignments in the `globally' non-distributive
quantal logic of the CH theory over its Boolean subalgebras in
that mutually compatible history propositions dwelling in the
latter take truth values in a Heyting algebra truth space which is
`larger' (although still distributive!) than the Boolean binary
alternative ${\mathbf 2}$ of classical (hence realist!)
set-logic\footnote{Only this observation could prompt one to look
for some sort of `quantum set theory'---a quantal extension of the Boolean
calculus of classical sets in the topos $\mathbf{Set}$ of constant
sets so as to account for the way quanta (represented by some kind
of `quantum sets') actually combine with each other. The upshot 
of such an endeavor would be the development of a corresponding  
`{\em quantum topology}' for small-scale spacetime structure 
\cite{isham89, isham91, df91}---to 
which we will return, in a bit more detail, in section \ref{algebran}. 
The general 
quantum set theory project was originally conceived by von Neumann \cite{vneum} and
has been significantly developed over the years along Grassmann and Clifford 
algebraic lines by Finkelstein
and coworkers. For the latest word from that research front, refer
to \cite{df96, sel98}. Of course, another possibility would be to 
formulate directly a `quantum topos'---a universe which would be a quantum 
version of $\mathbf{Set}$ thus it would provide a natural habitat for quantum 
sets. The search for such a quantum topos structure 
has been very broad and diverse \cite{bor1, bor2, nawaz, mul, mul1, mul2, sel91, rap, sel98, rap3, 
ces}, thus we will touch it only peripherally, and 
from a CH-theoretic point of view, in sections 
\ref{algebran} and \ref{affinities}.}.

We may summarize (c) and (d) in a geometrical sense by saying that
the non-distributive quantal logic of the CH theory is `warped' or
`curved' relative to its classical (Boolean) sublogics
\cite{rawse,rap3} and, internally in its topos
${\mathcal{T}}_{CH}$, it is locally intuitionistic---certainly not
two-valued, but still distributive. This intimate logico-geometric
interplay is allowed by the very essence of topos theory which is
widely known by now to unify logic and geometry at a deep level
\cite{lawv,maclane,rap3}.

\end{itemize}

In the next section we impart a topology of a special kind to the
presheaf objects in ${\mathcal{T}}_{CH}$ and subsequently we
sheafify them. The resulting category is the topos of {\em
sheaves} of sets over consistent-histories, thus we can expose
topological traits of the CH theory as well as qualify the
variable sets in ${\mathcal{T}}_{CH}$ to ones being {\em
continuously} variable in the sense of Lawvere \cite{lawv}.

\section{\Large\bf Topologizing and sheafifying consistent-histories}\label{sheafn}

In this section we get our hands dirty and become a bit more
technical than before, although we still present everything at a
`physical level of rigor' always referring to the relevant
mathematics literature for technical intricacies and results, as
well as to the pivotal paper \cite{isham97} for more analytical
discussion of various constructions and facts about history
propositions that are going to be quoted and used below. We extend
the topos ${\mathcal{T}}_{CH}$ in \cite{isham97} by providing a
Vietoris-type of topology to the second-level history propositions
dwelling in its presheaf objects ${\rm Set}^{{\mathcal{B}}}$ by
using the notion of sets `trapped' in $\mathcal B$. Subsequently,
this `topologization of histories' will enable us to sheafify the
objects of ${\mathcal{T}}_{CH}$. 

\subsection{\large\bf The abstract Vietoris topology}

First, we give a short and watered down exposition of the abstract
Vietoris topology that one can give to the collection
$\mathfrak{C}(X)$ of closed subsets $C$ of a topological space
$X$. More details may be found in \cite{michael,beer}. So, let $X$
be a topological space. With respect to any {\em open} subset $U$
of $X$ one can define:

\begin{itemize}

\item ($i$) \underline{\bf The `nerve' of $U$ in $\mathfrak{C}(X)$:}

\begin{equation}\label{e2}
U^{\cap}_{\mathfrak{C}(X)}:=\{ C\in\mathfrak{C}(X):~ C\cap U\not=\emptyset\}
\end{equation}

\item($ii$) \underline{\bf The `member' of $U$ in $\mathfrak{C}(X)$:}

\begin{equation}\label{e3}
U^{\subseteq}_{\mathfrak{C}(X)}:=\{ C\in\mathfrak{C}(X):~ C\subseteq U\}
\end{equation}

\end{itemize}

With these definitions of nerve (\ref{e2}) and member (\ref{e3}),
the Vietoris topology on $\mathfrak{C}(X)$ is defined as the one
generated by all basic sets of the form
$U^{\cap}_{\mathfrak{C}(X)}$ and
$U^{\subseteq}_{\mathfrak{C}(X)}$\footnote{Strictly speaking, the
collection of all nerves and members provides a {\em sub-basis}
for the Vietoris topology on $\mathfrak{C}(X)$, not a generating
set ({\it ie}, a basis) proper \cite{michael, beer}.}. We readily
apply this abstract definition to consistent-histories next.

\subsection{\large\bf A Vietoris-type of topology for consistent-histories}

The seed for the idea to endow consistent-histories with a
Vietoris-like topology can again be found in \cite{isham97}. As
the principal motivation for considering the Vietoris
topologization of consistent-histories one may regard Isham's
observation that the collection of second-level semantic values
$\mathfrak{V} (<a,p>)$, relative to a chosen decoherence
functional $\dd$, at an object (stage of truth) $W_{0}$ in
$\mathcal{B}$ of the form

\begin{equation}\label{eq4}
\mathfrak{V}^{\dd}_{W_{0}}(<a,p>):=
\left\lbrace\begin{array}{rl}
\{ W\subseteq W_{0}:~W\in\mathcal{B}^{\dd}~{\rm and}~ a\in W\} ~&\mbox if~\dd(a,a)=p\cr
\emptyset ~~~~~~~~~~~~~~~~~~~~~~~~~~~~~~~~~~~~~~~~~~&\mbox {otherwise}
\end{array}\right.
\end{equation}

\noindent do not form a logic algebra since the right hand side of
the defining equation (\ref{eq4}) is not a sieve\footnote{To
convince oneself of this fact, refer to \cite{isham97}. We also
note that ${\mathcal{B}}^{\dd}$ in (\ref{eq4}) is the poset
category on non-trivial Boolean subalgebras of $\latt$ metrized by
the $\com$-valued probability measure $\dd$.}. The result then is
that one cannot identify the subobject classifier $\Omega$ in the
topos ${\mathcal{T}}_{CH}$ of presheaves of varying sets over
${\mathcal{B}}^{\dd}$ with the Heyting logic algebra of the
collection of all sieves localized at the truth stage $W_{0}$, as
we mentioned in (b)-(d) of the previous section. In fact, at first 
sight one feels that one cannot apply at all the theory of
varying sets \cite{lawv,isham97}, thus {\it a fortiori} one cannot
view ${\mathcal{T}}_{CH}$ as a topos of presheaves of sets varying
over ${\mathcal{B}}^{\dd}$, once the coarse-graining sieve
structure and the intuitionistic logic calculus of all such sieves
breaks down object-wise ({\it ie}, locally) in the base poset
category ${\mathcal{B}}^{\dd}$. On the other hand, Isham points out
that ``...{\it in itself this}\footnote{That is, losing the sieve
structure object-wise in $\mathcal B^{\dd}$.} {\it does not rule
out the use of} (\ref{eq4}), {\it but it implies that any logical
structure on the set of semantic values must be obtained in a way
that is different from our anticipated use of the topos of varying
sets ${\rm Set}^{{\mathcal{B}}}$. One possibility is...}'' to
exploit the notion of `trapped sets'. We do this now.

One may observe, as Isham did in \cite{isham97}, that second-level
semantic values of the sort defined by expression (\ref{eq4}) do
not form a logic algebra, because they do not close algebraically
under set-theoretic union\footnote{See (A.2) in appendix A of
\cite{isham97}.}. To actually close them one may define finite
sets $F$ of history propositions that are `trapped' in Boolean
subalgebras $W$ of $\latt$ in ${\mathcal{B}}^{\dd}$ which
coarse-grain a particular stage of truth $W_{0}$, as follows:

\begin{equation}\label{eq5}
\mathfrak{T}^{\dd}_{F}(W_{0}):= \{ W\subseteq
W_{0}:~W\in\mathcal{B}^{\dd}~{\rm and}~ F\cap W\not=\emptyset\}
\end{equation}

\noindent where, plainly from (\ref{eq4}), $\mathfrak{V}_{a}^{\dd}(W_{0})
=\mathfrak{T}^{\dd}_{F}(W_{0})|_{F=\{ a\}}$.

The interesting feature of such trapped sets is that although they
close under set-theoretic union, they do not under intersection.
To establish $\cap$-closure one can consider the nerves of a
collection $\mathcal F$ of finite sets $F$ of history propositions
in $\latt$ relative to the coarse-grainings of $W_{0}$ in
${\mathcal{B}}^{\dd}$, as defined below:

\begin{equation}\label{eq6}
\begin{array}{l}
\mathfrak{T}^{\dd}_{{\mathcal{F}}=\{ F_{1},F_{2}\cdots
F_{n}\}}(W_{0}):= \{ W\subseteq W_{0}:~W\in\mathcal{B}^{\dd}~{\rm
and}\cr (F_{1}\cap W\not=\emptyset~\&~ F_{2}\cap
W\not=\emptyset~\&\cdots\&~F_{n}\cap W\not=\emptyset)\}
\end{array}
\end{equation}

The alert reader may have directly noticed in connection with
(\ref{eq6}) that some kind of Vietoris topology could be imposed
on consistent-histories in view of this expression's striking
formal similarity with the nerve (\ref{e2}) and member (\ref{e3})
expressions defining the abstract Vietoris topology on
$\mathfrak{C}(X)$. This is indeed so: one may define a
Vietoris-type of topology on ${\mathcal{B}}^{\dd}$ in $\latt$ by
taking as sub-basis the collection of all sets of the form
$\mathfrak{T}^{\dd}_{F}(W)~(\forall W\in{\mathcal{B}}^{\dd})$ as
$F$ ranges over all finite subsets of $\latt$. Let us symbolize
this topology by $\mathcal V^{\dd}$. We have thus effectively topologized the
base poset category ${\mathcal{B}}^{\dd}$ and, as Isham remarks in
appendix A of \cite{isham97}, the topological space
$(\mathcal{B}^{\dd},\mathcal{V}^{\dd})$ may be regarded as the
truth or semantic value space for consistent-history
propositions---the range of valuations localized or contextualized
on the Boolean windows $W$ of $\latt$.

Of course, this Vietoris-type of topology $\mathcal{V}^{\dd}$
assigned on $\mathcal{B}^{\dd}$, although it is not the same
Heyting logic algebra as in the case of the collection of all
coarse-graining sieves on $W_{0}$ which characterizes the
neorealist topos $\mathcal{T}_{CH}$ of varying sets proper in
\cite{isham97}, it still qualifies as a perfectly legitimate
example of an abstract (open set) topology---a complete
distributive lattice (of open subsets of the topological space
$(\mathcal{B}^{\dd},\mathcal{V}^{\dd})$) commonly known as a {\em
locale} \cite{maclane}\footnote{Interestingly enough, and from a
logic perspective, a complete distributive lattice is also known
as a complete Heyting algebra \cite{maclane}. The
consistent-history topos $\mathcal{T}_{CH}$ is, topologically
speaking, `locally localic'; while, logically speaking, its
internal logic is `locally Heyting' ({\it ie}, neorealist)
\cite{maclane,rap3,isham97}.}. Thus, by distorting a bit one's
point of view, one can still think (perhaps in an oblique sense)
of $\mathcal{T}_{CH}\equiv{\rm Set}^{\mathcal{B}^{\dd}}$ as the
topos of presheaves of sets varying over the poset category
$\mathcal{B}^{\dd}$---with this base space now having been
suitably topologized by $\mathcal{V}^{\dd}$. However, for the sake
of consistency, accuracy and clarity we must define presheaves of
sets over trapped sets of history propositions

\begin{equation}\label{eq7}
\mathbf{Tre}:~(\mathcal{B}^{\dd},\mathcal{V}^{\dd})\equiv\mathfrak{L}_{CH}\longrightarrow{\rm Set}
\end{equation}

\noindent in complete analogy to the presheaf objects in $\mathcal{T}_{CH}$

\begin{equation}\label{eq8}
\mathbf{Pre}:~(\mathcal{B}^{\dd})\longrightarrow{\rm Set}
\end{equation}

\noindent where in (\ref{eq7}) the base topological space
$(\mathcal{B}^{\dd},\mathcal{V}^{\dd})$ is identified with the
aforementioned locale $\mathfrak{L}_{CH}$ of its open
subsets\footnote{Note also in connection with (\ref{eq7}) that a
nickname for `presheaves over trapped sets' can be `tresheaves',
hence the symbol $\mathbf{Tre}$.}. The collection of the
presheaves $\mathbf{Tre}$ over the locale $\mathfrak{L}_{CH}$ is
another example of an abstract topos structure \cite{maclane},
which we may symbolize by $\tropos$\footnote{Quite reasonably, we
think, this `topos of tresheaves' may be coined `tropos' ({\it
gk.} for `manner' or `idiosyncracy'). This seems to be a suitable
name for the topos in focus in view of the peculiar character of
the unusual Vietoris topology carried by trapped sets in its
tresheaf objects.}. $\tropos$, in complete analogy with
$\mathcal{T}_{CH}$, may be thought of as universe of sets varying
over the locale $\mathfrak{L}_{CH}$ \cite{lawv,maclane}, hence its
internal logic too is neorealist in the sense of Isham
\cite{isham97}\footnote{In other words, the non-existence of
global sections of valuations ({\it ie}, the localization of
semantic values or truth) in $\mathcal{T}_{CH}$ carries through to
$\tropos$ so that localization maps {\it \`a la} (\ref{e1}) occur
in the latter's tresheaves, although, as noted above, in our tropos the
subobject classifier is a complete Heyting algebra or locale
different from the $\Omega$ of the coarse-graining sieves in
(\ref{e1}).}.

The bonus from working in $\tropos$ rather than in
$\mathcal{T}_{CH}$ is that having a sound background topological
space $\mathfrak{L}_{CH}$ over which sets vary in the tropos, we
can {\em sheafify} its presheaf objects relative to the Vietoris
topology $\mathcal{V}^{\dd}$, that is to say, we can promote the
contravariant functors ({\it ie}, presheaves
\cite{bredon,maclane,mallios,rap2}) in (\ref{eq7}) to `local
homeomorphisms' ({\it ie}, sheaves
\cite{bredon,maclane,mallios,rap2}) between the base topological
space $\mathfrak{L}_{CH}$ and the fiber or stalk space ${\rm
Set}$. We do this in the next subsection.

\subsection{Sheafifying consistent-histories}

Our sheafification of the presheaves in (\ref{eq7}) will be rather
swift. The procedure is quite a standard one and can be found in more
detail in \cite{bredon,maclane,mallios,rap3}. First we present the
general case of presheaves of functions over a topological space
$X$, then we particularize it to our case of tresheaves of sets
over the locale $\mathfrak{L}_{CH}$ in $\tropos$.

Initially, we note that presheaf maps such as the ones in
(\ref{eq7}) and (\ref{eq8}) are assignments to each open subset
$U$ in a topological space $X$ of function-like
objects\footnote{For the $\mathbf{Tre}$s and $\mathbf{Pre}$s in
(\ref{eq7}) and (\ref{eq8}) the objects assigned are sets in the
$\subseteq$-poset category ${\rm Set}$.} of the form
$S:~U\rightarrow S(U)$, and to each pair $(U,V)$ of open subsets
in $X$ nested by (strict) inclusion ({\it ie}, $U\subset V$) of a so-called
restriction map $\rho_{UV}:~S(V)\rightarrow S(U)$, subject to the
following conditions:

\begin{itemize}

\item (a) Identity: $\rho_{UU}={\rm Id}$.

\item (b) Composition: $\rho_{UV}\circ\rho_{VW}=\rho_{UW}~(U\subset V\subset W)$.\footnote{From
this rather standard definition of presheaves one can see clearly
why they are called contravariant functors: the direction of
$\subset$-arrows in the base or source poset category $X$ ({\it
ie}, when the topological space $X$ is regarded as the locale of
its open subsets) is reversed in the target poset category by the
presheaf maps.}

\end{itemize}

\noindent Thus, as it was mentioned above, this general definition
of presheaves prompts one to think of them as a collection of
functions on the open subsets of a topological space equipped with
restriction maps between them when their open set domains in $X$
are nested by inclusion\footnote{As it was mentioned before, for
the tresheaves in $\tropos$ these functions-like objects are just
structureless sets in ${\rm Set}$, but let us present the general
case first.}. In fact, one can construct a topological space
$\mathcal{S}$---the so-called {\em sheaf space}
\cite{bredon,maclane,mallios,rap2,malrap,rap3}---starting from a
presheaf of functions on a given topological base space $X$. Let
us recall briefly this well known construction which is commonly
known as {\em sheafification}.

First, for every open subset $U$ in the background topological
space $X$ define the so-called ``sections' selection map''
$\sigma_{U}$ from the set of presheaf functions $S(U)$ to a family
$\Gamma$ of continuous functions on $U$:
$\sigma_{U}:~S(U)\rightarrow\Gamma(U,\mathcal{S})$\footnote{$\sigma_{U}$
is assumed to commute with the $\rho$s in (a) and (b) above.}. The
elements of $\Gamma(U,\mathcal{S})$ are called the continuous
sections of $\mathcal{S}$ over $U$.

Second, define point-wise in $X$ ({\it ie}, for all $x\in X$) the
{\em stalks} (or fibers) $\mathcal{S}_{x}$ of the sheaf space
$\mathcal{S}$ as direct or inductive limits\footnote{Refer to \cite{maclane} 
for a definition of inductive systems of maps and their direct limits.} 
of the $S(U)$ presheaf maps above in the following way:

\begin{equation}\label{eq9}
\mathcal{S}_{x}:={\underrightarrow{\lim}}_{U\in\mathbf{B}(x)}\{ S(U):~ x\in
U\}\equiv\bigcup\{ S(U):~x\in U\} /\tilde{x}
\end{equation}

\noindent where $\tilde{x}$ is the following equivalence relation
between the functions in the $S(U)$s:

\begin{equation}\label{eq10}
f\tilde{x}g\Leftrightarrow \rho_{W,U\cap V}(f)=\rho_{W,U\cap V}(g),~(f\in S(U), g\in S(V))
\end{equation}

\noindent for some open neighborhood $W$ of $x$ in the `nerve' of
$U$ and $V$ ({\it ie}, for $W\subset U\cap V$)\footnote{In (\ref{eq9}), it 
is supposed that $U$ varies over a basis $\mathbf{B}(x)$ of 
open neighborhoods of $x$, while the maps in $S(U)$ constitute an inductive 
system of maps.}. As a
non-topologized set, the sheaf space is the disjoint union or
direct sum of its stalks:
$\mathcal{S}=\bigcup_{x}\mathcal{S}_{x}$.

Third, we topologize $\mathcal{S}$ as follows: define the {\em
germ of} $f$ {\em at} $x$, with $f\in S(U)$ and $x\in U$, to be
the $\tilde{x}$-equivalence class of $f$, and symbolize it by
$[f]_{x}$. Then, as a basis for the topology on ${\mathcal{S}}$ we
take the following family of open subsets:

\begin{equation}\label{eq11}
\mathfrak{B}[\mathcal{S}(X)]:=\{ (x,[f]_{x}):~x\in U\}
\end{equation}

\noindent A continuous section in $\Gamma(U,\mathcal{S})$ can then
be defined relative to this basis as:

\begin{equation}\label{eq12}
\sigma_{U}(f)(x)=[f]_{x}~(x\in U)
\end{equation}

\noindent and it is plain to see that the germs of $\mathcal{S}$'s
continuous sections dwell in its stalks, that is to say,
$[f]_{x}\in\mathcal{S}_{x}$. In fact, in this construction of the
topology on the sheaf space $\mathcal{S}$ relative to the base
topological space $X$, one can easily verify that the function
$\pi :~\mathcal{S}\rightarrow X$---called `the projection of the
sheaf space on the base space'---is a {\em local homeomorphism} 
\cite{bredon, maclane, mallios, mallios1, rap2}
acting on the basic open sets of $\mathfrak{B}$ in (\ref{eq11})
as:

\begin{equation}\label{eq13}
\pi(x,[f]_{x})=x
\end{equation}

\noindent By a {\em sheaf} one understands in general such a {\em
local homeomorphism}\footnote{Another, perhaps physically more intuitive, way
to say this is that the base topological space $X$ and the
overlying sheaf space $\mathcal{S}$ are locally ({\it ie},
$U$-wise in $X$) topologically equivalent or indistinguishable
and the sheaf $\pi$ implements this equivalence or indistinguishability 
\cite{rap2, malrap}. I wish to thank Tasos Mallios for bringing 
to my attention this (very physically-minded) definition of a sheaf 
originally due to Lazard \cite{cartan, mallios1}.}.

The stalks $\mathcal{S}_{x}$ of the sheaf carry the discrete
topology\footnote{This is another way of saying, as above, that as
a non-topologized set
$\mathcal{S}=\bigoplus_{x\in X}\mathcal{S}_{x}$.}, but as a topological
space proper it is generated by the germs of its continuous
sections $[f]_{x}$ inhabiting these very stalks. This is a well
known {\it clich\'e} in sheaf theory, namely, that {\em a sheaf
is} (generated by the germs of) {\em its continuous sections}
\cite{bredon, maclane, mallios, mallios1}.

So this is how a sheaf $\mathcal{S}$ arises from or is generated
by a presheaf $S$ on a topological space $X$. In fact, one can go
the other way around and note that the maps
$U\rightarrow\Gamma(U,\mathcal{S})$ constitute a presheaf that
satisfies the following `collation' properties:

\begin{itemize}

\item (a) If $U$ is covered by a family $\{ U_{i}\}$ of open subsets ({\it ie},
$U=\bigcup_{i}U_{i}$) and $s_{1}$, $s_{2}$ are sections in
$\Gamma(U,\mathcal{S})$ such that $s_{1}|_{U_{i}}=s_{2}|_{U_{i}}$
($\forall i$), then $s_{1}=s_{2}$.

\item (b) Let $\{ U_{i}\}$ be as above. If
$s_{i}\in\Gamma(U_{i},\mathcal{S})$ satisfy $s_{i}|_{U_{i}\cap
U_{j}}=s_{j}|_{U_{i}\cap U_{j}}$ ($\forall i,j$), then there is an
element $s\in\Gamma(U,\mathcal{S})$ such that $s|_{U_{i}}=s_{i}$
for each $i$.

\end{itemize}

\noindent Supposing that the presheaf $S:~U\rightarrow S(U)$,
subject to the usual $\rho$-restrictions as before, satisfies these two glueing
properties, one can show that the selection maps $\sigma_{U}$ are
in fact isomorphisms. That is to say, any presheaf satisfying (a)
and (b) above can be obtained as the presheaf of continuous
sections of a sheaf. Thus a sheaf may be reconstructed from its
presheaf of sections---the aforementioned {\it clich\'e}
vindicated.

As a matter of fact, these two procedures opposite to each other,
namely, sheafification of a (complete) presheaf\footnote{See
\cite{maclane,mallios, mallios1} for a definition of complete
presheaves.} and `pre-sheafification' from the continuous sections
of a (spatial) sheaf\footnote{Again, see \cite{maclane, mallios,
mallios1} for a definition of spatial sheaves.} are functors
adjoint to each other\footnote{See \cite{maclane} for a definition
of adjoint functors.} denoted by $\mathbf{S}$ and
$\mathbf{\Gamma}$, respectively \cite{mallios, mallios1}.

This exposition of the general sheafification procedure suffices
for our intention to sheafify the particular tresheaves of sets in
(\ref{eq7}). To this end, we make the following identifications:

\begin{itemize}

\item (i) The base topological space in our case is the locale
$\mathfrak{L}_{CH}$ of open subsets of the Vietoris topological
space $(\mathcal{B}^{\dd},\mathcal{V}^{\dd})$ of trapped sets.

\item (ii) The target category ({\it ie}, the range of the
presheaf maps $\mathbf{Tre}(U);~U\in\mathfrak{L}_{CH}$) is ${\rm
Set}$---the poset category of structureless sets ordered by
set-theoretic inclusion.

\item (iii) The sheaf space we will call $\mathfrak{S}_{CH}$, while
the sheaves resulting from the sheafification functor $\mathbf{S}$
on the $\mathbf{Tre}$s, $\mathfrak{S}_{CH}(\mathfrak{L}_{CH})$.

\item (iv) The basic open sets generating the topology on
$\mathcal{S}_{CH}$ are of the form $(a,[s]_{a})$, where $a$ is
just a singleton trapped set in $\mathfrak{L}_{CH}$ (arguably, of
point-like or `atomic'\footnote{Not necessarily an atomic 
proposition in the universal ortholattice $\latt$.} character!) 
and $[s]_{a}$ the fiber over it consisting
of the $\tilde{a}$-equivalence classes of sets in ${\rm Set}$
relative to the atomic history proposition $a$, much like
(\ref{eq9}) expressed in the general case of functions over 
the point-sets of $X$ rather than structureless 
sets\footnote{The sets belonging to the equivalence class
$[s]_{a}$ are `extensionally equal' ({\it ie}, with respect to
set-theoretic equality).}. In connection with (\ref{eq9}), we also note that 
the basic sets covering $a$ are taken, of course, from nerves and members 
in the sub-basis of $(\mathcal{B}^{\dd},\mathcal{V}^{\dd})$ that trap 
$a$ ({\it ie}, $\mathcal{T}^{\dd}_{\mathcal{F}=\{ a\} })$. 

\end{itemize}

Applying the aforementioned sheafification functor $\mathbf{S}$,
we have thus effectively obtained sheaves of sets varying {\em
continuously} \cite{lawv} over the Vietoris-topologized poset
category $\mathcal{B}^{\dd}$ without making use of the latter's
coarse-graining poset structure and its associated local
sieve-valued logical semantics. As a result, the collection
$\tropos^{\sigma}:=\{\mathfrak{S}_{CH}(\mathfrak{L}_{CH})\}$\footnote{The
superscript `$\sigma$' over $\tropos$ indicating `sheafification' of
the latter's tresheaf objects to sheaves.} of sheaves of sets over
$\mathfrak{L}_{CH}$ and sheaf morphisms between them is a topos
whose internal logic is inevitably intuitionistic \cite{maclane},
but not identical to the neorealist internal logic proper of 
$\mathcal{T}_{CH}$ in \cite{isham97}, as we contended earlier.

This concludes our presentation of sheafifying consistent-history
propositions in $\latt$, thus extending \cite{isham97} at the base
front. The next `reasonable' thing that one could do is to endow
the stalks of the $\mathfrak{S}_{CH}(\mathfrak{L}_{CH})$s in
$\tropos^{\sigma}$ with more algebraic structure, thus further extend
Isham's topos $\mathcal{T}_{CH}$ in \cite{isham97} even at the
stalk front. This is what we do in the next section. Before we do that, in 
the next subsection we discuss the physicality of the Vietoris topology 
on trapped sets and the sheafification process associated with it. 

\section{\large Algebraizing the stalks: the quantum causal histories topos}\label{algebran}

In the present section we extend Isham's work \cite{isham97} at the stalk
front as mentioned earlier by assigning more algebraic structure to the 
stalks of the sheaf objects $\mathfrak{S}_{CH}(\mathfrak{L}_{CH})$ of the 
tropos $\tropos^{\sigma}$, which stalks have so far been assumed 
to be occupied by structureless 
sets. Again, the procedure is quite a standard one: all that one 
has to make sure is, loosely speaking, that the additional algebraic
 structure employed in the fibers is compatible with or respects, locally at least, the `horizontal' 
continuity of the base topological space---its local topology so to speak---as it may, to preserve 
the sheaf structure\footnote{Another way to say this is that the extra algebraic operations defined 
stalk-wise in the  sheaves under focus should be continuous.}. Again, as we did for the 
sheafification of presheaves in the previous section, first we describe briefly the 
general procedure, which one may call {\em sheaf-algebraization} and can be found in more detail in 
\cite{bredon, maclane, mallios, mallios1}, then we specify the algebraic structures added to be the 
finite dimensional incidence Rota algebras modeling {\em quantum causal sets} (qausets) in 
\cite{rap1, malrap, rap3, ces}. Subsequently, we define the aforementioned Quantum Causal Histories 
Topos (QCHT) to be $\tropos^{\sigma\alpha}$\footnote{The superscript `$\alpha$' added to 
$\tropos^{\sigma}$ indicating now the Rota-algebraization of the 
set-inhabited stalks of the tropos' sheaf objects $\mathfrak{S}_{CH}(\mathfrak{L}_{CH})$. For the 
latter we also write $\mathfrak{S}^{\alpha}(\loc_{CH})$.} and we discuss briefly its affinities with 
Markopoulou's quantum causal histories scenario for quantum spacetime structure and gravity 
advocated in \cite{marko}. We also find some suggestive similarities with the curved finitary 
spacetime sheaves of qausets 
proposed in \cite{malrap} as a locally finite, causal and quantal model of (the kinematics of) the 
ever elusive Lorentzian quantum gravity, as well as with this model's non-commutative or quantum 
topological traits detected in \cite{rap3}.

\subsection{Rota-algebraizing the $\mathbf{\mathfrak{S}_{CH}}$s}

So, first we present the general sheaf-algebraization procedure {\it \`a la} 
Mallios \cite{mallios, mallios1}: the additional algebraic structures most commonly given to the 
stalks of a sheaf of structureless sets are $\com$-algebras or modules over such algebras. 
The most elementary example is that of a sheaf of (abelian) groups ({\it ie}, a group sheaf) 
$\mathcal{G}$ on a topological space $X$ whose stalks $\mathcal{G}_{x}$ are groups so that the 
(commutative) group operation, usually denoted by `$+$', is continuous in the following sense: 
defining the `fiber product' $\circ$ to be

\begin{equation}\label{eq14}
\mathcal{G}\times_{X}\mathcal{G}:=\{(g,g^{'})\in\mathcal{G}\times\mathcal{G}:~
\pi(g)=\pi(g^{'})\}\equiv\mathcal{G}\circ\mathcal{G}
\end{equation}

\noindent the map

\begin{equation}\label{eq15}
\mathcal{G}\circ\mathcal{G}\ni (g,g^{'})\mapsto g+g^{'}\in\mathcal{G}_{x}
\subseteq\mathcal{G}~(\pi(g)=\pi(g^{'})=x\in X)
\end{equation}

\noindent is continuous. Moreover, one can prove that the unary operation of inverting the group 
elements stalk-wise ({\it ie}, $g\mapsto g^{-1}\equiv -g$, $g\in\mathcal{G}_{x}$), hence of 
subtracting elements fiber-wise ({\it ie}, $g^{'}_{x}+g^{-1}_{x}\equiv g^{'}_{x}-g_{x}$), are also 
continuous in the manner above. Concomitantly, the group's neutral element $0$ ({\it ie}, 
$0_{x}=g_{x}+g^{-1}_{x}$, $\forall g_{x}\in\mathcal{G}_{x},~\forall x\in X$) is defined to be a 
global continuous section of $\mathcal{G}$.

Similarly to the definition of abelian group sheaves $\mathcal{G}$, one can define (unital) ring 
sheaves $\mathcal{R}$, $k$-algebra sheaves $\mathcal{A}$ ($k=\R$ or $\com$), as well as sheaves 
$\mathcal{M}$ of modules over such $k$-algebras ($k$-module sheaves)\footnote{Not insisting that the 
algebraic product is 
necessarily commutative. In fact, we will see shortly that the particular 
sheaf-algebraization of interest to us here will employ non-commutative 
rings and non-abelian $\com$-algebras.} by appropriately making sure that the extra structures 
imposed are continuous stalk-wise in the respective sheaves\footnote{For instance, the rings' 
multiplication unit $1$ defines a continuous global section of $\mathcal{R}$, while $k$-scalar 
multiplication is continuous in $\mathcal{A}$.}.

Particularizing the general sheaf-algebraization technique above to our case of interest,  
we assume that the stalks of the sheaf objects $\mathfrak{S}(\mathfrak{L}_{CH})$ in the topos 
$\tropos^{\sigma\alpha}$ are occupied not by sets, but by finite dimensional non-abelian incidence 
Rota $\com$-algebras $\amg$ \cite{zap1,rapzap1,rapzap2} representing qausets \cite{rap1,malrap}. The 
resulting structures are sheaves $\mathfrak{S}^{\alpha}_{CH}(\loc_{CH})$ of qausets over the 
Vietoris-topologized trapped sets of consistent-history propositions---in brief, `sheaves of 
consistent-histories of qausets'\footnote{In fact, since the incidence Rota algebras $\amg$ modeling 
qausets are $\Z$-graded $\com$-modules of 
`discrete differentials' over their commutative subalgebras $\amg^{0}$ of point-like `stationaries' 
\cite{rapzap1, rap1, malrap, rapzap2}, their sheaves are $\mathcal{M}$-sheaves in the sense 
above. This observation, namely, 
that the $\mathcal{M}$-sheaves of $\amg$s support discrete differential 
calculi and a discrete 
Riemannian geometry {\it \`a la} Dimakis {\it et al.} \cite{dim2, dim1} over consistent-histories, 
could prompt one to apply in this direction 
some very general, but powerful, concepts, results and techniques 
from Mallios' 
Abstract Differential Geometry on Vector and Algebra Sheaves \cite{mallios, mallios1} to the more 
conrete task of applying CH-theoretic ideas to quantum 
gravity. Such a possibility is roughly sketched in the subsection 4.3 
below, as well as in the 
concluding section.}. The topos structure having as objects 
these sheaves and as arrows sheaf morphisms between them is called the 
`{\em Quantum Causal Histories Topos}' and is abbreviated by the name's initials (QCHT). Like the 
general paradigm of a topos of sheaves of rings or algebras 
over a topological space $X$ or a locale $\loc$ can be interpreted as 
a mathematical universe of continuously variable rings or algebras varying with respect 
to the background `parameter space' $X$ or $\mathfrak{L}$ \cite{lambek, bell, maclane}, so the 
qausets inhabiting the stalks of the sheaves in the QCHT may be viewed as variable objects varying 
(continuously) relative to the Vietoris topology carried by the base poset category 
$\mathcal{B}^{\dd}$ in 
the universal consistent-histories orthoalgebra $\latt$. 

We strongly feel that the ultimate challenge for physics is  to find a  
plausible dynamics that `quantifies' this topos-variability of qausets, 
so as 
to qualify the topos-theoretic perspective to some kind of algebraic quantum 
gravity model proper \cite{malrap, rap3}. From this perspective one may perhaps 
get a clearer view of the supposedly central role that the CH theory plays 
in our quest for a cogent quantum theory of gravity. Unfortunately, in this 
paper we will not go as far as to give an explicit dynamics for qausets in their 
QCHT. Rather, we are going to content ourselves with drawing close connections 
between the QCHT and 
two recently proposed models of the kinematics of a finitary ({\it ie}, locally finite), causal and 
quantal version of (Lorentzian) gravity in \cite{marko} and \cite{malrap}. If anything, 
these connections will give us hints of how to develop further, and hopefully in the immediate 
future, the research program of
applying ideas from the CH theory to the problem of quantum gravity by sheaf and topos-theoretic 
means.

\subsection{Affinities between the QCHT and quantum causal histories}

Our brief comparison of the QCHT and Markopoulou's quantum causal histories scenario in \cite{marko} 
is centered around the observation that in the latter the base 
poset category on whose objects (vertices) finite dimensional Hilbert spaces $\mathcal{H}$ (of the 
same dimensionality)---realms in which states of a 
quantum system of a finite number of `degrees of freedom' (presumably, 
spacetime) are supposed to live---are 
localized, is taken to be a causal set (causet) $\vec{P}$ in the sense of Sorkin {\it et al.} 
\cite{bomb87, sor1, sor2, sor4, sor5, sor6}\footnote{ In 
a nutshell, a causet is a locally finite poset.}. From a finite number of Hilbert spaces soldered on 
a finite number of acausally (or `space-like') 
separated event-vertices in $\vec{P}$, 
tensor-product `compound' Hilbert spaces were then formed as befits the 
$\mathcal{H}$-representation theory of the CH formalism \cite{isham94}; moreover, unitary maps 
(modeling transitions) between such tensor product spaces were defined in a way that respects the 
reflexivity, antisymmetry and transitivity 
properties of the base poset $\vec{P}$\footnote{That is to say, in a way 
that respects the causal topology of the causet base space $\vec{P}$ 
\cite{malrap, rap3}.}. Thus, quantum causal histories and 
their unitary transformation theory were born and were held as sound 
unifications of the basic ideas of causet theory and the CH theory. The important thing to notice in 
this scenario is that the abstract temporal support 
on which consistent-histories are localized and relative to which they 
are supposed to vary, is provided by the finitary poset $\vec{P}$---needless to point out, as 
Markopoulou already did, that the collection of all `causal future sieves' based on any vertex of 
$\vec{P}$ form a Heyting algebra or locale.

Similarly, in our QCHT the abstract temporal background on which qausets are localized and with 
respect to which they are supposed to vary, is also a poset, 
namely, the complete distributive lattice $\loc_{CH}$, so that 
the QCHT itself is `locally localic' (or logically speaking, `neorealist'). 
The similarity 
becomes even more prominent if one decides to consider finite dimensional 
Hilbert space representations associated with the finite dimensional 
incidence Rota algebras dwelling in the stalks of the sheaf objects 
$\mathfrak{S}^{\alpha}_{CH}(\loc_{CH})$ in the QCHT\footnote{Such 
finite dimensional Hilbert space $\mathcal{H}$ matrix representations were studied in 
\cite{zap1}.}. The bundles associated with the $\mathfrak{S}^{\alpha}_{CH}(\loc_{CH})$s
are then such finite dimensional $\mathcal{H}$-vector sheaves 
in the sense of Mallios $\cite{mallios}$\footnote{See \cite{mallios} 
for a general definition of {\em associated sheaves} to vector, 
algebra and, more importantly, principal $\mathcal{G}$-sheaves. In 
the case of the $\mathcal{H}$-vector sheaves associated with 
the $\mathfrak{S}^{\alpha}_{CH}(\loc_{CH})$s in the QCHT, their sections 
represent generalized quantum states.}. Unitary-like transitions between 
the $\mathcal{H}$-stalks of these associated $\mathcal{H}$-sheaves 
are then induced 
by {\em geometric morphisms} on the QCHT\footnote{One may recall that, 
in general, with 
any bijective lattice morphism $f:~\loc\rightarrow\loc^{'}$ 
between two locales (which $f$ is, in effect, a homeomorphism between 
these two abstract pointless topological spaces), 
there is associated a pair of adjoint functors---the 
so-called `pushout' $f_{*}:~\mathbf{Sh}(\loc)
\rightarrow\mathbf{Sh}(\loc^{'})$ and `pullback' $f^{*}:~\mathbf{Sh}
(\loc^{'})\rightarrow\mathbf{Sh}(\loc)$---between the respective categories 
or topoi of sheaves (of any algebraic structures) over them \cite{maclane}. 
Within the particular QCHT, such functor pairs $(f_{*},f^{*})$---commonly known as 
geometric morphisms---are induced by elements $f$ of 
the group ${\rm Aut}(\loc_{CH})$ of automorphisms of $\loc_{CH}$.}.

On the other hand, however, there is {\it prima facie} a 
significant obstacle in carrying further this analogy between the QCHT 
and Markopoulou's quantum causal histories. If one decides to make 
use of the whole tensor product panoply of the CH theory underlying 
({\it ie}, providing the base space for) the sheaf obects $\mathfrak{S}^{\alpha}_{CH}
(\loc_{CH})$ of the QCHT, one is bound to 
encounter the following rather subtle technical difficulty. With the 
vector $\mathcal{H}$-sheaves associated to the $\mathfrak{S}^{\alpha}_{CH}
(\loc_{CH})$s a rather undesirable `{\em stalk-collapse phenomenon}' 
is observed whereby sections of two distinct $\mathcal{H}$-stalks over two distinct `atomic' history 
propositions ({\it ie}, $\mathcal{H}_{a_{1}}$ and 
$\mathcal{H}_{a_{2}}$) `merge' or `collapse' into a single stalk (now in the 
tensor product sheaf $\bigotimes_{i}\mathcal{H}_{i}$ over $\loc_{CH}$) when 
the underlying propositions tensor combine with each other as $a_{1}\otimes 
a_{2}$ \cite{isham94}. Since in the quantal logic of the CH theory $\otimes$ 
represents the phenomenon of {\em quantum entanglement} or {\em quantum 
coherence}, one may infer from the aforementioned stalk-collapse phenomenon 
that the usual `classical' tensor product structure is rather inadequate 
for representing the purely quantum behavior of entanglement, at 
least in a sheaf-theoretic context. Indeed, 
one may get a stronger feeling for the pathological character of  
this stalk-collapse if one assumes that the associated 
consistent-histories $\mathcal{H}$-sheaves are soldered on the 
points of a classical continuous spacetime manifold $M$ (or even 
on the pointless locale $\loc_{M}$ of open subsets about them, but 
this example is not as clear as in the case of sheaves 
on the pointed $M$)\footnote{Such 
an assumption to use the spacetime manifold $M$ as a base space would suit, 
for instance, a 
possible {\em continuous spacetime consistent-histories} theory \cite{isham98, sav1} with an eye 
towards applying ideas from the CH theory to the classical and,  
hopefully, to the ever elusive quantum theory of gravity \cite{sav2, sav3}.}. 

In such a hypothetical model of consistent-histories of qausets varying continuously 
over a background (possibly curved) spacetime continuum $M$, it is easy to 
see (at least from a more heuristic and  physical point of view) why the tensor 
product $\otimes$ and the `classical' definition of a sheaf
do not seem to go hand in hand: when one considers the tensor product of two distinct stalks in a 
vector sheaf like the associated ${\cal{H}}(M)$\footnote{Now, generalized quantum states of qausets 
are represented by $\mathcal{H}(M)$'s  
continuous sections.}, 
as when one combines two distinct quanta in the usual quantum theory, the two stalks `collapse' to a 
tensor product 
stalk over a single spacetime point-event of the classical base spacetime manifold $M$. This 
phenomenon is characteristic in both classical and quantum field 
theories where, when we coherently combine or entangle systems by tensor multiplication, 
their spacetime coordinates combine by identification. ``{\em This mathematical practice expresses a 
certain physical practice: to learn the time, we do not look at the system but at the sun (or 
nowdays) at the laboratory clock, both prominent parts of the episystem}'' \cite{df96}, and it 
should be emphasized that the episystem is always regarded as being classical\footnote{Here, the 
classical base spacetime manifold $M$.} in the sense of Bohr. 
All in all, this stalk-collapse pathology is begging for a radical revision of
`classical' tensor product $\mathcal{H}$-sheaves\footnote{Or even 
sheaves of tensor products of general Banach spaces.} over classical topological spaces in the 
sense that we should search for a new `quantum tensor product' 
structure $\otimes_{q}$\footnote{This symbol should not be confused 
with the one commonly used for the $q$-deformed group 
product of Hopf algebras and related quantum groups.} that soundly 
represents quantum entanglement and at the same time it evades the 
stalk-collapse observed 
in classical tensor product sheaves over classical pointed topological spaces 
(or even over their pointless locales)\footnote{Inevitably, this quantal version of the classical 
tensor product structure is expected to be accompanied by a quantum revision of 
`classical' sheaves and of the classical spacetime manifold 
topology on which these are defined (see the following paragraph and 
subsection). I   
wish to thank Chris Isham for a timely exchange on precisely such a possibility of a new `{\em 
quantum tensor product}' structure.}. 

Interestingly enough, 
and closely related to the quest for the $\otimes_{q}$ above, current 
researchers in quantum logic proper as well as in non-commutative or quantal 
generalizations of classical topological spaces (locales), are also looking for a similar quantal 
tensor product-like structure, which is non-commutative but distributes over the `$qor$' connective 
of the usual quantum logic \cite{birkhoff}, as it may,  
to replace the commutative, but not distributive over $qor$, `$qand$' operation of quantum 
logic\footnote{Jim Lambek in private communication.}. People have looked for 
such a $\otimes_{q}$ structure in Girard's linear logic \cite{gir} 
and in Yetter's non-commutative version of it \cite{yett}, as well as in 
Mulvey's quantale `$and$' or non-commutative $\cap$-like operation 
`\&' \cite{mul, mul1, mul2}\footnote{I wish to thank Steve Selesnick for bringing to my attention 
Girard's work and Mulvey's original quantale paper \cite{mul}.}.  
This author too has entertained the possibility for an analogous $\otimes_{q}$-connective 
in his endeavor to `{\em localize non-commutatively}' in the context of quantum gravity, that is to 
say, to develop a non-commutative topology and its associated 
sheaf or scheme theory \cite{voyst2, harts, voyst1} for the 
small-scale structure of spacetime \cite{rap3, ces} mainly motivated by the curved 
finitary spacetime sheaves (finsheaves) of qausets theme in \cite{malrap}. The search for the 
`right'  
$\otimes_{q}$ structure continues to stimulate quantum logicians and 
non-commutative topologists alike.

\subsection{Affinities between the QCHT and curved finsheaves of qausets}
 
The concluding words in the last subsection prompt us to present briefly 
some analogies between the QCHT, the curved finsheaves of qausets in 
\cite{malrap}, the latter's non-commutative topology suggested in \cite{rap3} 
and the related idea of a {\em quantum topos for quantum gravity} entertained in \cite{ces}. More 
details about all these ideas can be found in the corresponding citations and the references 
therein.

In \cite{malrap}, curved principal $\mathcal{G}$-finsheaves \cite{rap2} of qausets \cite{rap1} were 
proposed as reticular, causal and quantal replacements of the curved Lorentzian spacetime manifold 
$M$ of general relativity---classical gravity's kinematical structure. As base spaces for these 
finsheaves, causets $\vec{P}$ were assumed in a manner similar to the $\mathcal{H}$-localization 
spaces in the quantum causal histories scenario of Markopoulou that we briefly 
encountered above. These finsheaves were subsequently subjected to a 
`classicalization coarse-graining' procedure in the sense that an inverse system or net  
({\it ie}, coarse-graining poset category {\it \`a la} $\mathcal{B}$ in \cite{isham97}) consisting 
of finer-and-finer such finsheaves possessed at the limit 
of infinite localization or resolution or refinement (of spacetime into its point-events \cite{sor3, 
rap2}) a limit $\mathcal{G}$-sheaf isomorphic to the 
spin-Lorentzian principal fiber bundle of classical gravity\footnote{This 
is just a $\mathcal{G}$-bundle with structure group $SL(2,\com)$---the double cover of the 
orthochronous Lorentz local gauge group of general relativity.}. In turn, this inverse limit 
localization procedure was physically interpreted 
as Bohr's correspondence principle in a way originally proposed in the case 
of discrete quantum spacetime topologies modeled after finite dimensional 
Rota incidence algebras by Zapatrin and this author in \cite{rapzap1}.

Since, as it was also mentioned earlier, qausets are discrete differential 
manifolds {\it \`a la} Dimakis {\it et al.} \cite{dim1, dim2}, discrete $sl(2,\com)$-valued 
connections $\mathcal{D}$ were defined as $\mathcal{G}$-finsheaf morphisms by following closely 
Mallios' Abstract Differential Geometry on Vector Sheaves theory in \cite{mallios, mallios1}. The 
central point made in \cite{malrap} is that these $\mathcal{G}$-finsheaves admit no global 
$\mathcal{D}$-sections, so that they qualify as being `curved'. Subsequently, and in \cite{rap3, 
ces}, the idea was pitched 
to organize these $\mathcal{G}$-finsheaves into a topos-like structure 
$\mathbf{Sh}_{fcq}$\footnote{The topos of (f)initary, (c)ausal and (q)uantal sheaves of qausets.} 
which may be viewed as a locally finite, causal and quantal substitute for the `classical' topos 
$\mathbf{Sh}(M)$ of sheaves of sets over the 
spacetime continuum $M$\footnote{This $\mathbf{Sh}_{fcq}$ is one candidate 
for the quantum topos structure briefly alluded to in footnote 36.}. The  
aforementioned non-existence of global $\mathcal{D}$-sections of the reticular $\mathcal{G}$-sheaf 
objects of $\mathbf{Sh}_{fcq}$ is completely analogous to 
the non-existence of global valuations in the QCHT or in the presheaf topos $\mathcal{T}_{CH}$ of 
\cite{isham97}. The former entails a non-trivial curvature form 
on the $\mathcal{G}$-finsheaves. A resonable question one might ask 
is whether there is an analogous `{\em quantum logical curvature form}' on the 
$\mathfrak{S}^{\alpha}_{CH}(\loc_{CH})$ objects in the QCHT?\footnote{We will return to this 
question in the next section where we entertain 
the possibility of a cohomological classification of the $\mathcal{A}$-sheaf 
objects of the QCHT {\it \`a la} Mallios \cite{mallios, mallios1}.}. What 
is worth stressing at this point is that only topos theory allows for such a close 
logico-geometric interplay, that is to say, to speak to speak of a 
geometric spacetime curvature (gravity) in the $\mathcal{G}$-finsheaves 
of \cite{malrap} {\em and} of a sort of quantum logical or semantic 
curvature in the warped sheaves of the QCHT  \cite{lawv, maclane, sel91, rap3}.

As we mentioned in the previous subsection, the problematic tensor product stalk collapse in 
$\mathcal{H}$-vector sheaves, here too we mention a problem 
that may arise with 
the curved $\mathcal{G}$-finsheaves of qausets in \cite{malrap}. The finite 
dimensional vector $\mathcal{H}$-sheaves associated with them, whose sections 
represent quantum states of qausets, are supposed to carry a representation 
of the reticular spin-Lorentz structure group of the principal $\mathcal{G}$-finsheaves. If on top 
we would like to emulate the situation in the quantum 
causal histories approach discussed above thus wish to implement unitary transitions 
between the stalks of the associated $\mathcal{H}$-sheaves, we would soon run into significant 
problems, 
because {\em there are no finite dimensional unitary representations of the 
Lorentz group since it is non-compact}. Of course, one could resort to an 
`easy-way-out' by saying on the one hand that the reticular and quantal version of the 
spin-Lorentz structure group of the $\mathcal{G}$-finsheaves of qausets neither 
a continuous (Lie) nor even a `classical' group any more, and on the other that the continuous 
spacetime manifold together with the continuous group of its symmetries somehow `emerges' (as a 
macroscopic effect) from such quantal $\mathcal{G}$-finsheaf substrata and it does not have to be 
accounted for at 
quantum scales\footnote{See \cite{malrap} for 
more arguments about this.}. On the other hand, one could ultimately question the 
validity of unitarity 
in the quantum deep, since the latter is a non-local conception ({\it eg}, 
in non-relativistic quantum mechanics unitarity involves an integral over all 
space, while in quantum field theory, over all spacetime), but that would appear to kill quantum 
causal histories altogether.

We conclude this subsection by remarking on a possible `consistent-histories 
of non-commutative or quantum spacetime topologies' scenario in the QCHT. In \cite{rap3} 
it was argued that the non-abelian Rota incidence algebras modeling qausets 
are also finitistic non-commutative topologies suitable for a 
quantum theoresis of spacetime topology\footnote{Furthermore, in 
\cite{rap3} and subsequently in \cite{ces} it was conjectured 
that the topos $\mathbf{Sh}_{fcq}$ of the `non-commutative sheaves' 
mentioned in the previous paragraph 
may be the canonical example of a {\em quantum topos} structure 
in the same way that the collection of (commutative) sheaves over a 
locale is the canonical paradigm of a `classical' topos. 
See also \cite{bor1, bor2} for a similar, but technically much more 
sophisticated, conception 
of quantum topoi and the non-commutative topology/sheaf theory 
that they encode.}. Very recently, this gave birth 
to the related idea that spacetime topology can be regarded as a quantum observable 
of a foam-like nature \cite{rapzap2}. However, much earlier, and in the 
context of Rota-algebras and their $\mathcal{H}$ representations, not only 
spacetime topology had been conceived as being subject to some sort of quantum measurements and 
dynamical fluctuations \cite{grib}, but also that one could even formulate a histories theory for 
such quantum spacetime topology measurements and variations \cite{brezap}. Even 
more suggestive is the observation that for the formulation of a theory 
of {\em quantum topology} on the lattice of all topologies on a set 
$X$ of fixed finite cardinality (and canonical ({\it ie}, Hamiltonian) 
dynamical transitions between them) \cite{isham89}, Vietoris-type of topologies 
like the one we used here 
for topologizing and concomitantly sheafifying the CH theory may play a crucial role \cite{beer, 
isham91}. Thus 
the question arises: can we marry all these diverse ideas under the single QCHT roof?---a question 
generating a quest that is certainly worth pursuing further in the future. 

\section{\large Brief comparison with the Kochen-Specker topos and a future 
outlook}\label{affinities}

To the future project that closed the last section we would like to add 
and discuss briefly a couple more below.

The first project for the immediate future that we would like to suggest is to 
try to relate the two strikingly similar topoi of presheaves of sets that 
arise in connection with the semantic analysis of the quantal logic of the CH theory in 
\cite{isham97} ({\it ie}, the $\mathcal{T}_{CH}$ above) on the 
one hand, and on the other 
from considering similar valuation-localizations over the Boolean sublattices 
of a quantum lattice that result from viewing the famous Kochen-Specker (KS) theorem of quantum 
logic proper from a topos-theoretic perspective \cite{buttish1, buttish2, buttish3, 
rawse}\footnote{Let us call this topos `$\mathcal{T}_{KS}$'---`the Kochen-Specker topos'.}. 

It is immediately transparent from comparing the $\mathcal{T}_{CH}$ and 
$\mathcal{T}_{KS}$ topoi that both use a poset category of Boolean sublattices 
of the proposition ortholattices underlying their quantal amd quantum logics 
respectively as base spaces for semantic localizations, both toposes possess 
presheaf objects that do not admit global valuation-sections thus, as a result, both are `warped' 
relative to their classical Boolean subalgebras or sublogics
and have a neorealist ({\it ie}, Heyting) logic calculus as their internal 
contextualized logic\footnote{This `warped' or, geometrically speaking, 
curved  character of the quantal and quantum logics of the CH theory and the usual quantum theory 
respectively is to be contrasted against the `intrinsically flat' character of 
classical Boolean logic. From a sheaf-theoretic perspective, for example, it 
is well known that any Boolean algebra may be equivalently cast as the algebra 
of global sections of a sheaf of $\mathbf{2}$s over its Stone space, hence it 
is flat \cite{gold, bell, rawse}. In 
this way one may also understand the fact that in the intuitionistic (or 
neorealist!) topos $\mathbf{Sh}(X)$ of sheaves of continuously variable sets 
over a topological manifold $X$, $\mathbf{Set}$---the archetypal Boolean topos of constant sets 
whose subobject classifier is $\mathbf{2}$---arises as an instantaneous 
`snapshot' or localization of a geometric morphism kind of $\mathbf{Sh}(X)$ on $X$'s points or 
`instances' \cite{sel91}.}. Indeed, these remarkable similarities
call for a closer comparison between the quantal logic of the CH theory and 
the usual quantum logic in the illuminating light of sheaf and topos theory. 
At the same time, this unified topos perspective is even more formidable if one considers the 
significant differences on the interpretational side between the logic of the 
CH theory, whose propositions are in a strong sense `{\em diachronic}' ({\it ie}, about entire 
histories of quantum systems), and quantum logic proper whose propositions are 
well known to be `{\em synchronic}' or instantaneous ({\it ie}, at a single 
moment of time about the observable properties of quantum systems)---the dramatic differences 
between the (neo)realist philosophy 
supporting the CH theory and the operationalist one supporting the usual `Copenhagen' quantum theory 
aside.

The second project that we would like to bring forth is a possible infusion 
of differential geometric ideas into the CH theory with an eye towards 
applying the resulting structures to quantum gravity---a problem that 
consistent histories are expected to address sooner or later. As we 
mentioned in the previous section, we hope to apply quite straightforwardly 
concepts and results from Mallios' general and abstract ({\it ie}, axiomatic) 
treatment of the usual differential calculus on manifolds by means of vector 
and algebra sheaves to the sheaves $\mathfrak{S}_{CH}^{\alpha}(\loc_{CH})$ 
of consistent-histories of qausets and their associated $\mathcal{H}$-state sheaves 
in their QCHT. In particular, since it has been well established that the finite dimensional 
incidence Rota algebras modeling qausets are discrete differential manifolds supporting discrete 
differential calculi in the sense that a 
reticular version of the nilpotent K\"ahler-Cartan differential 
$\mathbf{d}$ \cite{degreech, iasc}  
has been shown to be operative on these algebra sheaves \cite{malrap}, this 
$\mathbf{d}$ may be able to trigger the start of a de Rham-like cohomology 
theory on the $\mathfrak{S}_{CH}^{\alpha}(\loc_{CH})$ objects in the QCHT---arguably the focal point 
in Mallios' {\it aufbau} of the usual $C^{\infty}$-differential 
geometric constructions in a more abstract, but powerful, sheaf-theoretic setting \cite{mallios1}. 
Related to the above, and as it was also briefly mentioned earlier, since the $\mathcal{G}$-
finsheaves of qausets are curved \cite{malrap}, one could also entertain the possibility of 
cohomologically classifying {\it \`a la} Mallios \cite{mallios, mallios1} the similarly warped 
algebra and vector sheaves of histories 
of qausets in their QCHT by means of a {\em characteristic curvature class} 
$\mathfrak{F}$, in a manner analogous to how it is usually done in the case of the locally trivial 
fiber bundles that are of interest to Yang-Mills theories and, possibly, to (quantum) gravity 
\cite{avis}. 

The possibility of bringing together such fiber bundle techniques and sheaf cohomology ideas from 
Mallios' work, and apply them to the quantum 
causal history sheaves $\mathfrak{S}_{CH}^{\alpha}$ of interest here, rests on the observation 
that the sheaf $\mathcal{S}$ of germs of continuous sections of a
locally trivial $k$-vector bundle $\mathbf{B}$ ($k=\com ,\R ,
\cdots$) has the property that there is a an open cover
$\mathcal{U}=\{ U_{i}\}$ of the base topological space $X$ such that
for each covering set $U_{i}$ in $\mathcal{U}$:

\begin{equation}
\Gamma(U_{i},\mathcal{S})\cong\bigoplus_{n}C^{0}(U_{i},k)
\end{equation}

\noindent where $C^{0}(U_{i},k)$ denotes the collection of continuous
functions from $U_{i}$ into $k$. It is well known that 
this sheaf preserves the action
of the bundle's \v{C}ech cocycles so that $\mathbf{B}$ may be
reconstructed from its sheaf of germs of sections, as well as 
from the algebraic structure of the set of sections which, in
turn, inherits the algebraic structure of the objects that one may
assume to inhabit the stalks of the sheaf $\mathcal{S}$\footnote{See section
\ref{algebran} and \cite{mallios, mallios1}.}---these two constructions being essentially
equivalent \cite{sel91, mallios, mallios1}. So, in view of the (quantum) logical character 
of the algebra sheaves $\mathfrak{S}^{\alpha}_{CH}(\loc_{CH})$ and the sheaf cohomology 
project anticipated above, one may reasonably ask: is there some kind of `quantum 
logical curvature' characteristic class $\mathfrak{F}$ classifying the 
$\mathfrak{S}_{CH}^{\alpha}(\loc_{CH})$ objects in the QCHT?---certainly a question worth pondering 
on. However, the possibility of such a geometrical characterization of 
`quantum logical' sheaves, such as our $\mathfrak{S}_{CH}^{\alpha}(\loc_{CH})$s, 
essentially by algebraic means ({\it ie}, via sheaf cohomology) \cite{mallios, 
mallios1}, hence also the possibility of 
bringing together, quite unexpectedly, quantum logic and quantum gravity 
concepts and constructions, we have only lately begun to fathom---being surely guided in our quests 
by the illuminating light of topos theory. Perhaps, such a potential conceptual unity
between quantum spacetime structure and dynamics on the one hand, and quantum logic on the other, 
achieved 
by applying sheaf and topos-theoretic ideas to the CH theory and quantum 
logic in particular, as well as to the general quantum 
gravity program \cite{buttish4, malrap, rap3}, will further vindicate Finkelstein's deep insight in 
\cite{df96} that ``{\em logics come from dynamics}'' \cite{ces}. 

\section*{Acknowledgments}

Chris Isham, Jim Lambek, Tasos Mallios, Chris Mulvey, Steve
Selesnick and Roman Zapatrin are all gratefully acknowledged for
numerous discussions on the general possibility of realizing Einstein's
vision about ``{\itshape an entirely algebraic description of
reality}'' \cite{einst56} along sheaf and topos-theoretic lines,
as well as for their moral and material support over the years. 
Special thanks to Chris Isham for bringing to my attention appendix 
A in \cite{isham97} where the alternative idea to use trapped sets 
and a Vietoris-type of topology on them, rather than sieves on 
the base poset category $\mathcal{B}^{\dd}$ in $\latt$, 
was originally conceived.  
Thanks are also due to the EU for the financial support in the
form of a generous Marie Curie postdoctoral research fellowship
held at Imperial College.

\end{document}